\documentclass[11pt,preprint,a4paper]{aastex}

\newcommand{\ms}{\mbox{m\,s$^{-1}$}}

\newcommand{\kms}{\mbox{km\,s$^{-1}$}}
\newcommand{\msun}{M$_{\odot}$}

\newcommand{\mjup}{M$_{\rm Jup}$}

\newcommand{\mterr}{M$_\oplus$}

\newcommand{\msini}{$m \sin i$}

\newcommand{\lrhk}{$\log$R$^{\prime}_{\mathrm{HK}}$}

\shortauthors{O'Toole {\it et~al.\/}}
\shorttitle{Low-mass Exoplanets}
\slugcomment{ApJ   Submitted:                 Accepted:             }
\received{}
\accepted{}
\begin{document}

\title{The Frequency of Low-mass Exoplanets}
\author{
S.~J.~O'Toole\altaffilmark{1,2},
H.~R.~A.~Jones\altaffilmark{2},
C.~G.~Tinney\altaffilmark{3},
R.~P.~Butler\altaffilmark{4},
G.~W.~Marcy\altaffilmark{5,6},
B.~Carter\altaffilmark{7},
J.~Bailey\altaffilmark{3},
R.~A.~Wittenmyer\altaffilmark{3}
}
\email{otoole@aao.gov.au}

\altaffiltext{1}{Anglo-Australian Observatory, PO Box 296, Epping 1710, Australia }
\altaffiltext{2}{Centre for Astrophysics Research, University of Hertfordshire, Hatfield, AL 10 9AB, UK }
\altaffiltext{3}{Department of Astrophysics, School of Physics, University of NSW, 2052, Australia }
\altaffiltext{4}{Department of Terrestrial Magnetism, Carnegie Institution of Washington, 5241 Broad Branch Road NW, Washington DC, USA 20015-1305}
\altaffiltext{5}{Department of Astronomy, University of California, Berkeley, CA USA 94720 }
\altaffiltext{6}{Department of Physics and Astronomy, San Francisco State University, San Francisco, CA, USA 94132 }
\altaffiltext{7}{Faculty of Sciences, University of Southern Queensland, Toowoomba, Queensland 4350, Australia }

\begin{abstract} We report first results from the Anglo-Australian
Telescope Rocky  Planet Search -- an intensive, high-precision Doppler
planet search targeting low-mass exoplanets in contiguous 48 night
observing blocks. On this run we targeted 24 bright,  nearby and
intrinsically stable Sun-like stars selected from the
Anglo-Australian Planet Search's main sample.  These observations have
already detected one low-mass planet reported elsewhere (HD\,16417b),
and here we reconfirm  the detection of HD\,4308b. Further, we have
Monte-Carlo simulated  the data from this run on a star-by-star basis
to produce robust  detection  constraints. These simulations
demonstrate clear differences in the exoplanet detectability functions
from star to star due to differences in sampling, data quality and
intrinsic stellar stability. They reinforce the importance of
star-by-star simulation when interpreting the data from Doppler planet
searches.  The simulations indicate that for some of our target stars
we are sensitive to close-orbiting planets as small as a few Earth
masses. The two low-mass planets present in our 24 star sample
indicate that the exoplanet minimum mass function at low masses  is
likely to be a flat $\alpha \sim -1$ (for $dN/dM \propto M^{\alpha}$)
and that between 15$\pm10\%$ (at $\alpha=-0.3$) and 48$\pm34\%$ (at
$\alpha=-1.3$) of stars host planets with orbital periods of less than
16 days and minimum masses greater than 3\,\mterr. 
\end{abstract}


\section{Introduction}
\label{sec:intro}

The planetary mass function is a key constraint which theories of star
and planet formation must be able to accurately predict, if they are
to be considered viable. While a consensus is emerging  that  planet
formation, in general, is probably dominated by the growth  of rocky
cores  via the accretion of dust particles and/or ices
\citep[i.e. the ``core accretion'' paradigm developed to explain the formation of the Solar System, e.g.][]{Pollack96},   the primary
alternative scenario of direct gravitational collapse remains a
plausible mechanism \citep[e.g.\ ][]{Boss2007}.

Three indirect lines of reasoning suggest that terrestrial-mass
planets could orbit at least a few percent of stars. The first planet
with a minimum mass of less than 10\,\mterr\ -- GJ\,876d -- was
identified just 3 years ago.  GJ\,876d  has a minimum (\msini) mass of
4.9\,\mterr\ and a probable mass of 7.5\,\mterr\ \citep{RLB05}.  In
the years since, a further five Doppler exoplanets have been announced
with minimum masses of less than 10\,\mterr\
\citep{Udry07,Mayor08}. And a further two low-mass exoplanets have
been revealed by gravitational microlensing searches
\citep{Beaulieu06,Bennett08}. So there can be little doubt that
exoplanets below 10\,\mterr\ do indeed exist, though whether such
planets are rocky ones (like the Earth) or icy ones (like Neptune and
Uranus) is yet to be unambiguously demonstrated.  Second, gas-giant
planets  exist around at least 6.5\% of nearby main sequence stars
\citep{marcy05} and planet formation theories suggest  that rocky
cores and embryos should accompany such Jovian planets
\citep{WS89,GLS04,KB06}. Finally, dusty protoplanetary disks are very
common around young  T Tauri stars, which suggests the ubiquity of the
building blocks of rocky planets \citep{HSC98,HLL01,MMH04}.

While not definitive, these observations  are suggestive of the
existence of rocky planets outside our Solar System. However, the
theory of rocky planet formation remains  less than solid. We do not
know what fraction of stars form rocky planets, nor how many of those
planets avoid dynamical ejection by the larger planets. Worse, we do
not know  if rocky planets with masses above 1\,\mterr\ quickly
accrete  gas and volatiles to become ice-giants like Neptune and
Uranus. This has been predicted by \citet{IL04} and \citet{GLS04}, and
would result in a planetary mass desert between 1 and  14\,\mterr, as
indeed we see in our Solar System. Theory does suggest that disks
with higher surface-mass density than the minimum-mass  solar nebula
are expected to produce numerous rocky planets  formed closer to their
parent star, which would make their Doppler detection more feasible
\citep{IL04}.  Moreover, planets of 1-15\,\mterr\ may retain water in
amounts that are comparable to the silicates and iron-peak elements,
resulting in a family of  ocean-planets \citep{LSS04}. 

What is clear, is that the detection of a statistically meaningful
sample of planets between 5-25\,\mterr, along with measurements of
their masses, radii, and orbits,  will be required to significantly
inform these diverse theories, and drive forward our detailed
understanding of how planets form.

\section{The Anglo-Australian Planet Search}
\label{sec:aaps}

The Anglo-Australian Planet Search (AAPS) is a long-term  radial
velocity program targeting the detection and parametrisation of
exoplanets.  The AAPS' main survey targets 250 southern Sun-like
stars, and has been in operation since 1998. To date the survey has
discovered 32 new  exoplanets orbiting stars in the main sample
\citep{AAPSI,AAPSIII,AAPSVII,AAPSXI,AAPSXIII,AAPSII,AAPSV,
AAPSIV,AAPSVI,AAPSVIII,AAPSXII,AAPSIX,AAPSX,AAPSXIV,AAPSXV}.  AAPS'
precision Doppler measurements are made with the UCLES echelle
spectrometer \citep{diego:90}.  An iodine absorption cell provides
wavelength calibration from 5000 to 6200\,\AA.  The spectrograph
point-spread function and wavelength calibration are derived from the
iodine absorption lines embedded on every pixel of the spectrum by the
cell \citep{val:95,BMW96}.  This observing and analysis  system has
demonstrated long term precisions of 3\,\ms\ for late-F, G, and
early-K dwarfs brighter than V=7.5  \citep{AAPSXI,AAPSII}.

The UCLES dispersion through a 1\arcsec\ slit gives an effective
resolution of $\lambda/\Delta\lambda$=60000. A precision of 1\,\ms\
corresponds  to displacements of 0.5\,millipixel at our detector, and
this, then, is the level below which we  aim to hold the sum of all
systematic uncertainties. Many subtle uncertainty  issues enter at the
level of 0.5 millipixel, such as determining  the wavelength scale to
9 significant digits, establishing the  PSF shape at 1 part in 1000
(which varies dramatically with  wavelength and seeing), measuring the
charge transfer efficiency of the CCD, and calibrating the non-linear
response of the quantum efficiency of the CCD. In addition, we
specifically  treat the Earth's telluric H$_{\mathrm{2}}$O and
O$_{\mathrm{2}}$ lines that contaminate our spectra at levels below
0.5\% intensity, and we eliminate all ghosts and defects from  the
spectrometer and CCD, on a pixel by pixel basis. We also  employ a
fully relativistic Doppler correction to the barycenter of the Solar
System to remove the effect of the Earth's orbital motion and
rotational spin.  

Our data processing procedures follow  those described by
\citet{BMW96,AAPSII,BWM06} and \citet{AAPSXI}. All data taken by the
AAPS to date have  been reprocessed through our continuously upgraded
analysis system. The results presented in this paper arise from the
current version  of that pipeline.

\section{The Anglo-Australian Rocky Planet Search}
\label{sec:rocky}

The detection of very low-mass exoplanets by the AAPS and other
Doppler programs within the last 4-5 years has in large part been due
to the dramatic improvements achieved in the measurement  precisions
these searches achieve.  These have  improved to such an extent that
it is now clear that noise sources {\em intrinsic} to the parent star
themselves are the limiting factor for very low-mass exoplanet
detection. Characterisation of these noise sources (jitter, convective
granulation and asteroseismological p-mode oscillations) has become an
important focus of Doppler planet detection. A few obvious
modifications to current observing strategies have emerged -- (1)
target low-mass stars; (2) target chromospherically inactive and
slowly rotating stars; (3) target high-gravity stars (where p-mode
oscillations are minimised) and (4) extend the observations of stars
over several p-mode fundamental periods, so that asteroseismological
noise is averaged over.

The Anglo-Australian Rocky Planet Search specifically seeks to focus
on the last three of these observing strategies, in an effort to push
to the lowest possible detection limits achievable with our system.
That is, it focusses on the  observation of bright stars known to be
relatively stable to radial velocity jitter (e.g.  Wright 2005).  We
ensure that all observations of  each target are of sufficient length
to average over the dominant asteroseismological period for each
target star. For this purpose, we employed the algorithms of
\citet{OTJ08} to derive the maximum asteroseismological beat period
($P_{\mathrm{max}}$) for each star, and then ensure that observations
are {\em at least} this long. Observations of our Rocky Planet Search
targets, which typically spanned 15-20 minutes, easily achieved this
goal.

Finally, and most importantly, observations are carried out over long,
contiguous observing runs of 48 nights.  These long observing blocks
are critical, as they allow us to integrate up Fourier power  and
suppress sidebands in the window function over many  periods. The
first of these Rocky Planet Search observing blocks was scheduled on
the 3.9m Anglo-Australian Telescope (AAT) from 2007 Jan 10 to 2007 Feb
26, spanning two full bright lunations and the dark and grey time in
between.

The power of this intensive observing approach is most simply and
dramatically demonstrated by our detection of the short-period planet
HD\,16417b \citep{AAPSXVI} from 24 epochs obtained on that run,  and
the {\em non-detection} of the same planet in a similar quantity of
similar quality data spread over a 2 year period.  This bears out the
simulations performed by other investigators  --  that to detect
planets with amplitudes of 1-to-3 times that  of the net measurement
uncertainty, it is necessary to acquire observations over many
contiguous periods \citep[see for example,][]{NCL05}.

\subsection{Sample Selection}
\label{sec:sample}

The stars chosen for Rocky Planet Search observation on this first run
were a subset  of the AAPS main sample \citep{AAPSVI} that are bright
(V$<$6.7), inactive (\lrhk$ < -4.8$) and in the right ascension range
0-17$^h$ with southerly declinations, suitable for observation in
January and February at the Anglo-Australian Telescope.  The sample
all had primary asteroseismological period, $P_{\mathrm{max}}$, less
than 800s. It includes 24 F, G and K  stars.  No metallicity criteria
were used in this selection. Stars  with known stellar companions
within 2 arcsec are removed  from the observing list  as it is
operationally difficult to get an uncontaminated spectrum  of a star
with a nearby companion.  One star which satisfied these criteria
(HD\,75289) has a known  exoplanet in a 3.5\,d orbit
\citep{UMN00,BWM06} of such high-amplitude (K=54\,\ms) that it would
have significantly complicated the possibility of detecting a low-mass
planet. As a result it was not included in the observed
sample. Otherwise, stars were not rejected on the basis of having
known exoplanets. HD\,4308 was selected for inclusion in this sample
when the observing program was first proposed in 2005, and was
retained for observation even after a low-amplitude planet (K=4\,\ms)
was announced in a 15.1\,d orbit by \citet{UMB06}.

The stars observed, and their properties, are summarised in  Table
\ref{tab:targets}.  All of these stars have been included in
large-scale studies  of nearby solar type stars. \citet{Nordstrom04}
included  them in a magnitude-limited, kinematically unbiased study of
16682 nearby F and G dwarf stars. \citet{VF05} included them in a
study of the stellar properties for 1040 F, G and K stars observed for
the Anglo-Australian,  Lick and Keck planet search programmes. Valenti
\& Fischer  used high signal-to-noise ratio echelle spectra
(originally taken as  radial velocity template spectra) and spectral
synthesis to derive effective temperatures, surface gravities and
metallicities whereas \citet{Nordstrom04} used Str{\"o}mgren
photometry and the infrared flux calibration of \citet{Alonso96}. Both
studies use Hipparcos parallaxes to obtain luminosities and make
comparisons with different  theoretical isochrones to derive stellar
parameters. To determine stellar masses and ages Valenti \& Fischer
use Yonsei-Yale isochrones \citep{Demarque04}, while
\citet{Nordstrom04}  use Padova isochrones
\citep{Giradi00,Salasnich00}. Both sets of derived  parameters are
consistent to within the  uncertainties of these studies.

\subsection{Observations}
\label{sec:obs}

\begin{figure*}
  \begin{center}
    \includegraphics[clip=true,width=15cm]{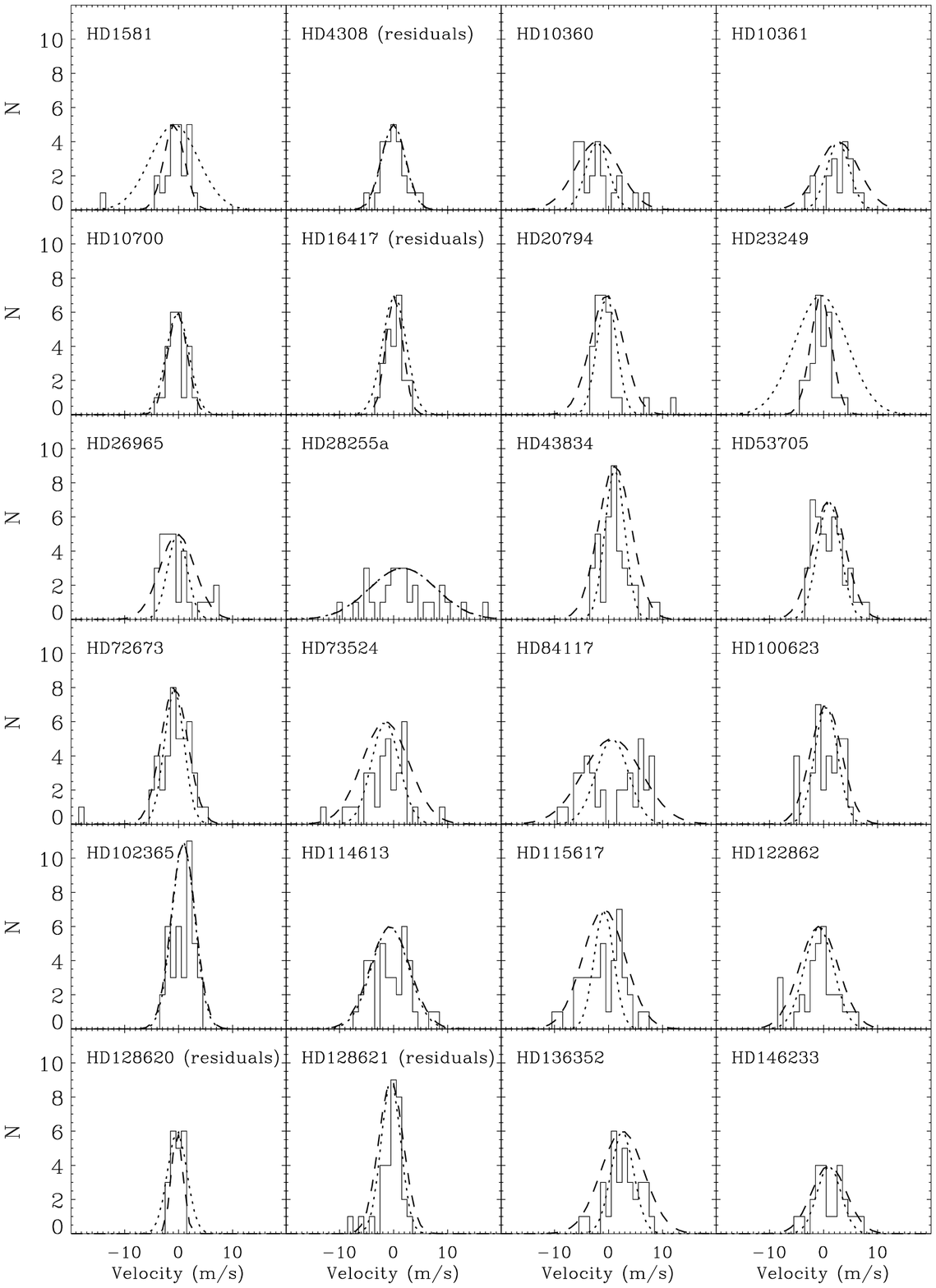}
    \caption{Histograms of measured Doppler velocities for each of our
      24 targets. Also plotted are Gaussians with FWHM equal to the
      stellar jitter (dotted) and the velocity scatter
      (dashed). Histograms of the residuals are shown for the two
      objects with known planets (HD\,4308 and HD\,16417), while the
      binary system $\alpha$ Cen (HD\,128620 and HD\,128621) have had
      a linear trend subtracted.}
    \label{fig:vels}
  \end{center}
\end{figure*}

The observing strategy on each night was straightforward -- to observe
every star in the target list for at least 15 minutes (or as long as
is required to obtain S/N$\approx$400 per spectral pixel given the
available transparency and seeing conditions) on each of the 48 nights
of observing.  The number of observations actually obtained for each
of the 24 targets is given in Table \ref{tab:obs} and ranges from 18
to 44, with a median of 31. The second column in this table provides
the root-mean-square (RMS) for each target about the mean velocity,
with four exceptions -- the values given for the two stars with
planets (HD\,16417 and HD\,4308 -- see below) are the RMS of the
residuals after a best-fit planet was subtracted, and the values for
the two components of the binary system $\alpha$ Cen (HD\,128620 and
HD\,128621) have had linear trends subtracted. The ``Act. jitter''
column shows the activity jitter derived using an updated version
\citetext{Jason Wright, private communication} of the \citet{Wright05}
recipe, which is considered to produce jitter estimates good to
$\pm$50\%.  The ``Osc. jitter'' column gives the oscillation jitter
\citep[from ][]{OTJ08}  for the average exposure time used for each
target.  To present these results graphically, histograms of the
measured Doppler  velocities are shown in Figure \ref{fig:vels} for
each object (with  the plots for HD\,4308, HD\,16417, HD\,128620 \&
HD\,128621 showing residuals as described above).  Overplotted are
Gaussians with full-width at half-maximum (FWHM) equal to the RMS
(dashed line) and the activity jitter (dotted line) for each object.

\begin{figure}
  \begin{center}
    \includegraphics[clip=true,width=7.5cm]{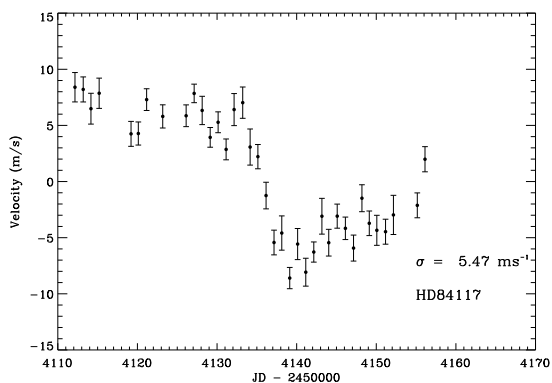}
    \caption{Data from Rocky Planet Search run for HD\,84117.}
    \label{fig:84117}
  \end{center}
\end{figure}

There is at least one case (HD\,84117) where the observed velocity
variability distribution appears to deviate significantly from a
Gaussian distribution. We therefore plot the actual velocities
obtained for this object in Fig. \ref{fig:84117}, which show evidence
for significant variability over the course of the 48n run, though
with no obvious periodicity, nor with an obviously Keplerian
shape. (No other star observed on this run  shows a similar
variability trend, indicating that the variability seen is not a
systematic effect of our measurement system).  Adding in additional
AAPS data of similar quality taken since 2005  Jan 30 indicates that
HD\,84117 has shown excess velocity variability since 2005 over that
expected from activity jitter alone (with an RMS of 4.7\,\ms). The
periodogram of HD\,84117 shows essentially no power at periods of less
than 40\,d, though a complex, broad power distribution is seen at
longer periods, i.e. longer than the time-span of the observations. We
can therefore fit the data with no compelling single
Keplerian.  HD\,84117 may either contain multiple planets, which will
require intensive observation to disentangle, or may be a star with an
unusual class of velocity variability.  We can nonetheless say with
confidence that it does not host a low-mass exoplanet in an orbit of
less than $\sim$30\,d.

Thirteen of the 24 stars have observed RMS values consistent with
their jitter estimates to within $\pm$50\% and all but two are
consistent to within a factor of two -- this is in line with
expectations if the jitter estimates have a 1-$\sigma$ uncertainty of
$\pm$50\%.  The two outliers in this comparison  are HD\,1581 (F9.5V)
and HD\,23429  (K0V), where the observed velocity variation is much
less than that predicted by the jitter model (by factors of  2.3 and
2.5, respectively), and HD\,115617 (G5V) where the level of
variability observed is larger than that predicted by the jitter model
(by a factor of 2.2).  The activity jitter estimate does not include
velocity variability due to asteroseismological oscillations (which is
in general much smaller than the activity jitter -- see Table
\ref{tab:obs}) and more importantly, does not include the effects of
convective granulation, which is (to date)  unparametrised for any
Doppler planet search stars.  What can be concluded is that at very
high velocity precisions,  even ``inactive'' stars remain velocity
variable at low levels.

\subsection{Planets in the Sample}
\label{sub:hd4308}

There are two objects in our target list that have detectable planets
orbiting them: HD\,4308  and HD\,16417. The latter was discovered  as
part of this observing campaign with P=17.24$\pm$0.01\,d,
$e$=0.20$\pm$0.09,  and \msini=22.1$\pm$2.0\,\mterr. It is presented
in a separate paper \citep{AAPSXVI}, to which we refer the reader for
further details. 

\begin{figure}
  \begin{center}
    \includegraphics[clip=true,width=8.0cm]{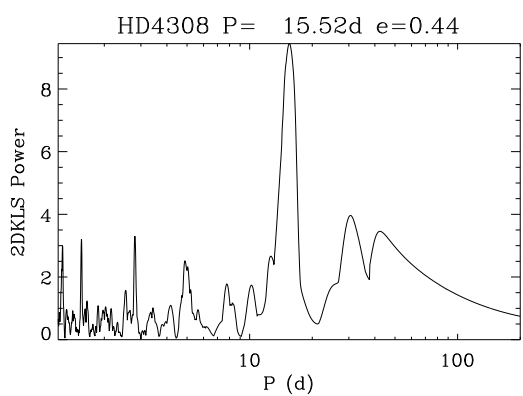}
    \caption{2DKLS Periodogram evaluated at $e=0.44$ determined for
      HD\,4308 from AAT Rocky Planet Search data alone.}
    \label{fig:hd4308power_mr}
  \end{center}
\end{figure}

\begin{figure}
  \begin{center}
    \includegraphics[clip=true,width=7.5cm]{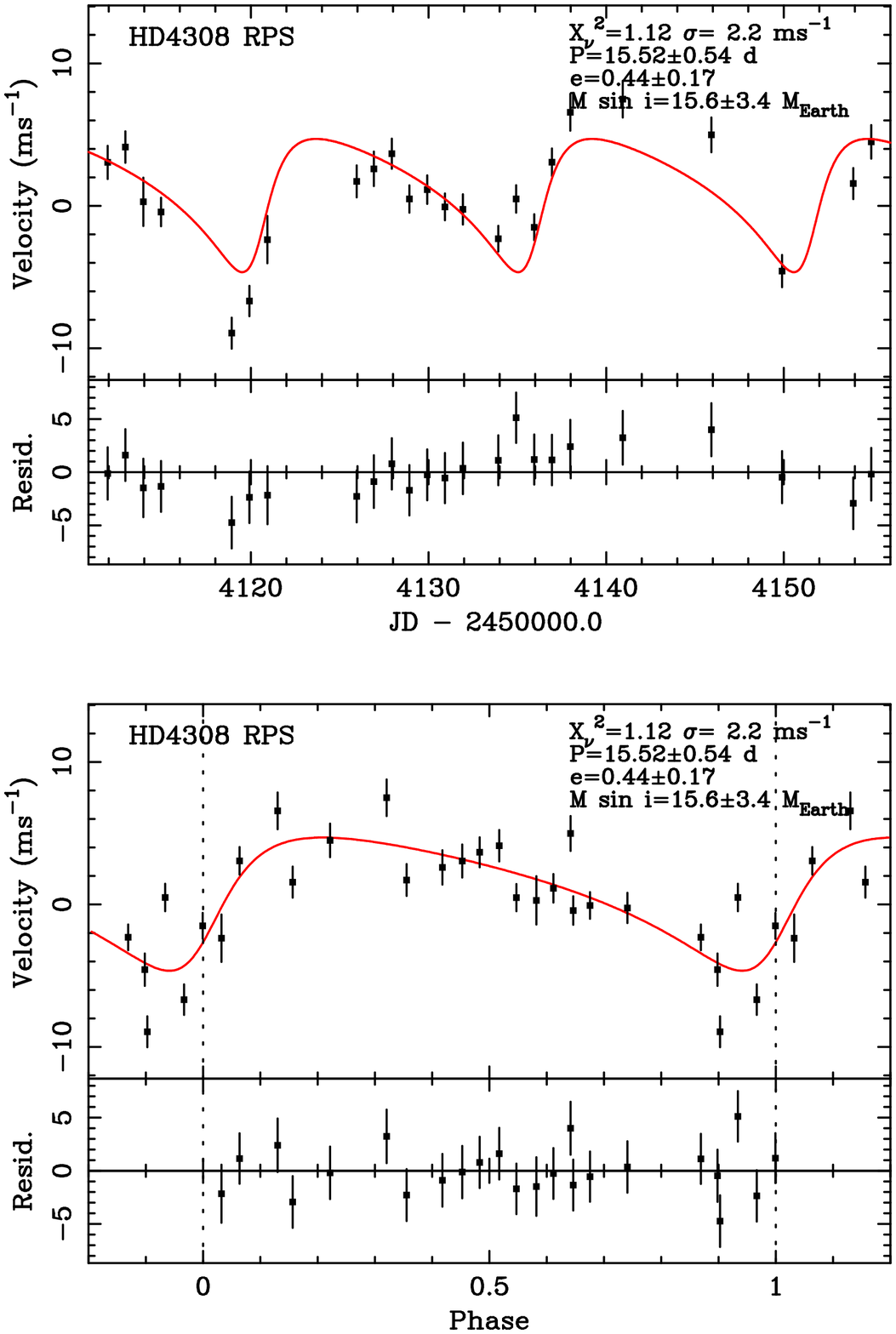}
    \caption{Keplerian fit to HD\,4308 data from Rocky Planet Search (RPS) run.}
    \label{fig:hd4308}
  \end{center}
\end{figure}

HD\,4308b was discovered by \citet{UMB06} and is a $\sim$\,13\,\mterr\
planet in a 15.6\,d orbit around a G5 dwarf. It was observed 25 times
during our campaign and the 15.6\,d period is clearly evident in the
2D Keplerian Lomb-Scargle (2DKLS)  periodogram \citep{AAPSXIV} shown
in Figure \ref{fig:hd4308power_mr}.  The data from this run alone are
indeed sufficient to a Keplerian fit which measures similar orbital
parameters (see Fig. \ref{fig:hd4308}) to the Udry et al.\ data. A
scrambled-velocities false-alarm-probability test \citep[FAP -
][]{marcy05} indicates that for 5000 trials, $\sim$16\% of scrambled
velocities for this data set give a reduced chi-squared better than
the solution.

\begin{figure}
  \begin{center}
    \includegraphics[clip=true,width=7.5cm]{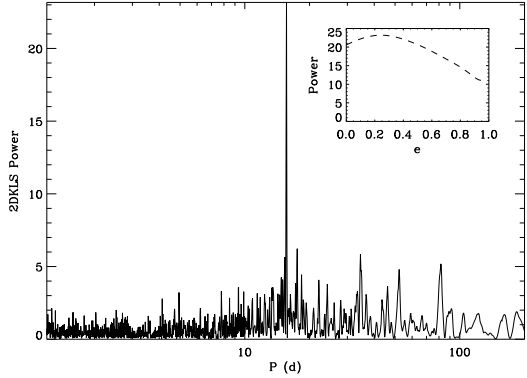}
    \caption{2DKLS periodogram determined at $e=0.24$ for HD\,4308   
    from AAT data (2005 Oct-2007 Nov, including Rocky Planet Search data) and HARPS data \citep{UMB06}. 
    The inset shows 2DKLS power as a function of eccentricity at constant period, 
    and demonstrates how weakly this data set (in common with most Doppler data sets) constrains eccentricity.}
    \label{fig:hd4308power_all}
  \end{center}
\end{figure}

To refine the orbital parameters of HD\,4308b, we have
combined the velocities published in \citet{UMB06} with our AAPS
velocities -- this includes both the observations obtained of HD\,4308
in the Rocky Planet Search run of 2007 Jan-Feb, and other AAPS
observations obtained with similar SNR between 2005 Oct 21 and 2007
Nov 27. These observations are listed in Table \ref{vel4308}.  The
2DKLS periodogram based on this combined dataset is shown in Figure
\ref{fig:hd4308power_all}. The inset in this figure shows the
corresponding  slice through the 2DKLS at constant period, as a
function of eccentricity, and  demonstrates how eccentricity is not
well constrained by this data set. 

\begin{figure}
  \begin{center}
    \includegraphics[clip=true,width=6.5cm]{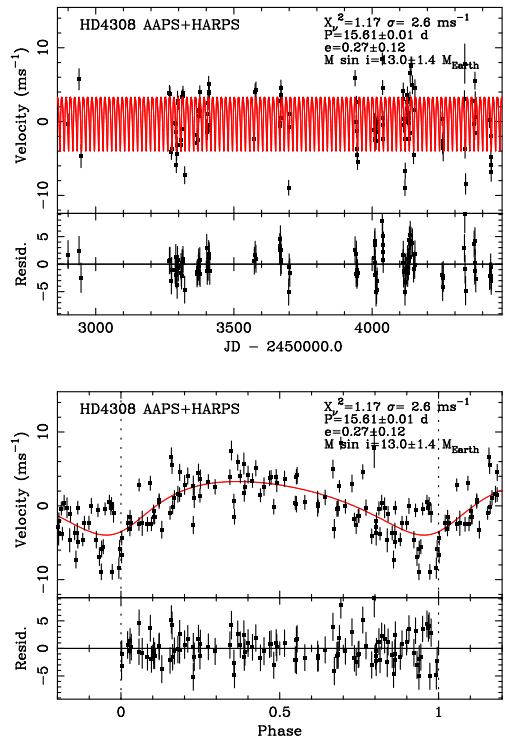}
    \caption{Velocities for HD\,4308 from the AAT (2005 Oct-2007 Nov, including Rocky Planet Search data) 
      and HARPS \citep{UMB06} plotted against time (upper panel) and phased at the best-fit
      orbital period (lower panel). The 
      line indicates the best-fit Keplerian to this combined data set, the parameters of which
      are listed in Table \ref{tab:hd4308}. }
    \label{fig:hd4308all}
  \end{center}
\end{figure}

In Figure \ref{fig:hd4308} we show our best fit to the HD4308 data
from the Rocky Planet Search run alone, while in Figure
\ref{fig:hd4308all} we show the combination of all high-SNR data from
the AAT and published HARPS data, and list the derived parameters in
Table \ref{tab:hd4308}. (The mass for HD\,4308 of
0.91$\pm$0.05\,\msun\  due to \citet{VF05} is adopted, as is a jitter
estimate for HD\,4308 of 2.17\,\ms).  We have determine a FAP
for this planet of $<$0.02\%, based again on 5000 scrambled velocity
datasets. Though the eccentricity derived from
the combined data set ($e=0.27\pm 0.12$) is higher than that published
by Udry et al., we note two points in relation to this; first, that it
has been recently demonstrated by several studies \citep{OTJ09,ST08}
that there is a systematic bias against measuring zero eccentricities
for low-amplitude planets; and second, that the uncertainty estimates
for eccentricity produced by the least-squares fitting of Keplerians
can seriously underestimate the true eccentricity uncertainty
represented by a data set \citep{OTJ09}. Given this, we do not believe
the differences between the Keplerian parameters for HD\,4308b
derived here and by Udry et al. are significant.

\begin{figure}
  \begin{center}
    \includegraphics[clip=true,width=6.5cm]{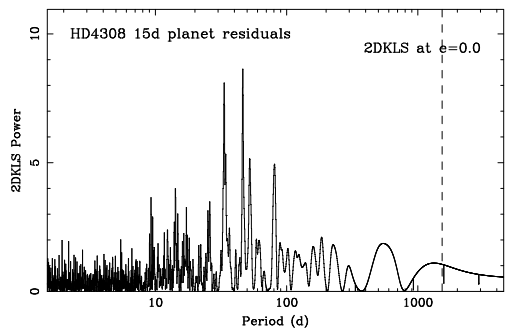}
    \caption{2DKLS at $e$=0.0 for the HD\,4308 velocities of Table \ref{vel4308} 
    after removal of the best fit Keplerian of Fig \ref{fig:hd4308all}. While there is some evidence for excess
    power at 30-80\,d periods, this is not currently sufficient to justify
    claiming the detection of a second planet. (The vertical dashed line indicates the
    time span of the observations used to make this periodogram.)}
    \label{fig:hd4308resid}
  \end{center}
\end{figure}

Finally, we note that the residuals to the Rocky Planet Search data
alone (Fig. \ref{fig:hd4308}) are suggestive of a further periodicity
in that data set at 30-40\,d. To examine the possibility of there
being a second planet present in this data, we have constructed the
2DKLS for the AAT and HARPS velocities with the Keplerian of
Fig. \ref{fig:hd4308all}  removed (see
Fig. \ref{fig:hd4308resid}). The result is suggestive of power at
period between 30-80\,d.  While potential planets at 32\,d and 48\,d
periods can be fitted to this data, they do not do so with a
significance that warrants a claim to have detected further planets in
this system.

\section{Simulations}
\label{sec:sims}
 
The biases inherent in planet search observing and analysis strategies
remain the largest single hurdle to a robust understanding of the
formation processes that have built the 300-odd exoplanets known to
date. Considerable work has been conducted on trying to eliminate
these underlying  biases from Doppler exoplanetary statistics
\citep[e.g.\
][]{OTJ09,2006AJ....132..177W,Cumming08,Cumming03,Cumming99,Cumming04}. 

To assess the selection functions delivered by the AAPS, we have been
working towards a detailed understanding of Doppler noise sources
intrinsic to stars \citep[e.g.\ ][]{OTJ08}. We then use detailed
object-by-object Monte-Carlo simulations \citep{OTJ09}, to explore the
biases introduced by these noise sources, when combined with  our
observational window functions. 

The procedure we employ is to generate single Keplerians for a grid of
periods, eccentricities and semi-amplitudes, which are then sampled at
the actual observation time-stamps and which have noise added to them
in line with the  actual measurement uncertainties and stellar noise
appropriate for each epoch.  These simulated observations are then
subjected to an automated detection process, enabling us to determine
the range of period, planet mass and orbital eccentricity to which
each data set is sensitive. The simulations are performed using the
Keter and Swinburne supercomputer facilities.  Details of the
simulations and detectability criteria are described in \citet{OTJ09}
-- though with the modification to the procedure in that paper (which
describes the analysis of purely artificial data), that the
simulations of {\em actual} data, used here, include added noise terms
due to intrinsic stellar velocity variability \citep{Wright05,OTJ08}.

To simulate the 48 nights of data obtained on this first Rocky Planet
Search run, we used a subset of the model grid from \citet{OTJ09} with
input periods on a logarithmic scale from $\log_{\mathrm{10}}P=0.3,
0.6, 0.9, 1.2$ (i.e. periods of between 2\,d and 16\,d);  input
eccentricities of 0.0, 0.1 \& 0.2;  and planet masses of 0.01, 0.02,
0.05, 0.1, 0.2 and 0.5\,\mjup. The eccentricity range simulated was
deliberatelylimited to the ``near-circular'' range of $e$=0.0-0.2,
since (a) the  majority of planets in such small  orbits
(i.e. $<$0.12\.AU) will have undergone significant tidal
circularisation, and (b) the vast majority of detected Doppler
exoplanets with periods of less than 16\,d have eccentricities, $e <
0.2$ \citep[e.g.][]{marcy05japan}. 
As our data is approximately evenly sampled (each star was
observed on average once per night at approximately the same time), we
have supplemented the above with additional simulations at the
same eccentricities and masses, but with non-integer periods of 2.8,
4.3, 5.6, 8.7, 11.2, 17.4 and 22.4 days. Because the
Keplerian function of a Doppler exoplanet is unable to determine
inclination angle to the line of sight, $i$, no attempt has been made
to deal with this in these simulations. As a result, whenever we refer
to ``mass'' for a simulated planet in the following discussion, we are
actually probing the \msini\ `` minimum mass'' of that planet. For
each (period,eccentricity,planet mass) point in this grid, 100
realisations are performed each with random noise added appropriate to
the combined impact of the uncertainty at that time-stamp and the
jitter and p-mode noise appropriate for that star (see Table
\ref{tab:obs}). Each star, then, is the subject of 21600 simulations. 

These simulations allow us to generate estimates of the detectability,
for each star, of putative planets at each  (mass, period,
eccentricity) point in the simulations.  In this context, we define
``detectability'' at a given set of orbital parameters as the number
of detections \citep[using the detection criteria of ][]{OTJ09}
divided  by the corresponding total number of realisations. We have
then integrated this detectability over eccentricity since (a)
detectability is approximately constant with eccentricity at the
values we have simulated \citep[see ][]{OTJ09}, and (b) we wish to
focus on our survey's sensitivity to planet mass and period. The
result is a set of surfaces indicating detectability as a function of
\msini\ and period. These surfaces are shown  in Fig. \ref{fig:detect}
for two examples  (HD\,4308 and HD\,53705),  as well as for the whole
sample of 24 objects. 

The specific examples shown for HD\,4308 and HD\,53705 display the
general features seen in the simulations of all the targets: first
that detectability is very high at periods longer than a few days and
at masses above 20\,\mterr; and second, that the details of the
detectability contours vary from object to object. The primary cause
for this is the different time and SNR sampling of the  observations
of each star.  This reinforces the importance of simulating
radial-velocity data on a star-by-star basis, rather than on a survey
sensitivity basis. We not that the structure in the surfaces at
$\sim$\,2 and $\sim$\,4 days is an effect of sampling; the
observations were taken on average once per night at approximately the
same time. This means that planets with periods that are integer
multiples of this have poorly constrained parameters and are therefore
harder to detect.

And finally, we must ask how many planets are detected from our actual
observations if we apply to them the same automated criteria used to
detect planets from the simulated data sets? Reassuringly, we find
that just two planets are detected (HD\,4308b and HD\,16417b, although
the former is perhaps marginal) -- none
of the other stars reveal planets that pass our automated detection
criteria.

\section{Discussion}

\begin{figure}
  \begin{center}
    \leavevmode
    \includegraphics[clip=true,width=9.0cm]{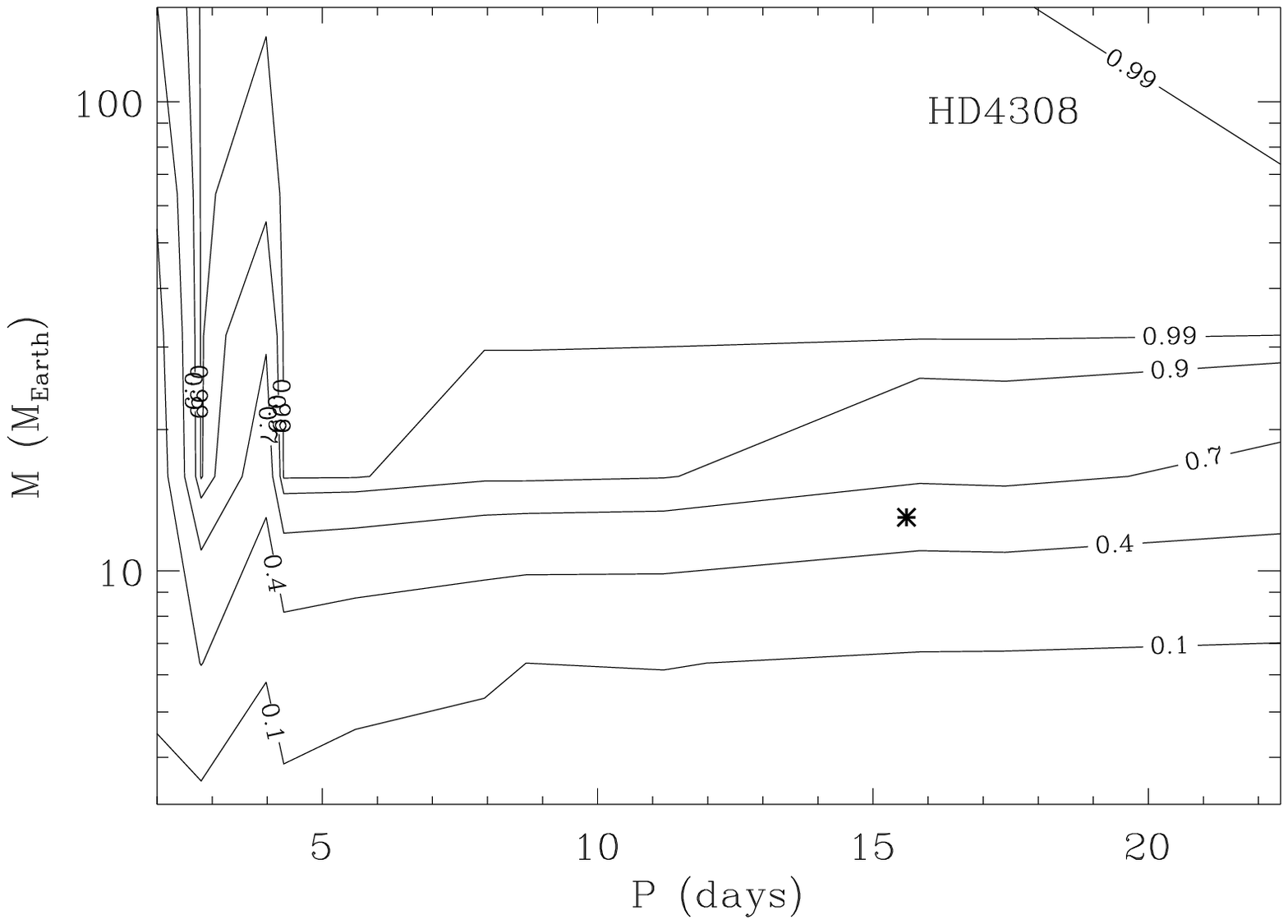}
    \includegraphics[clip=true,width=9.0cm]{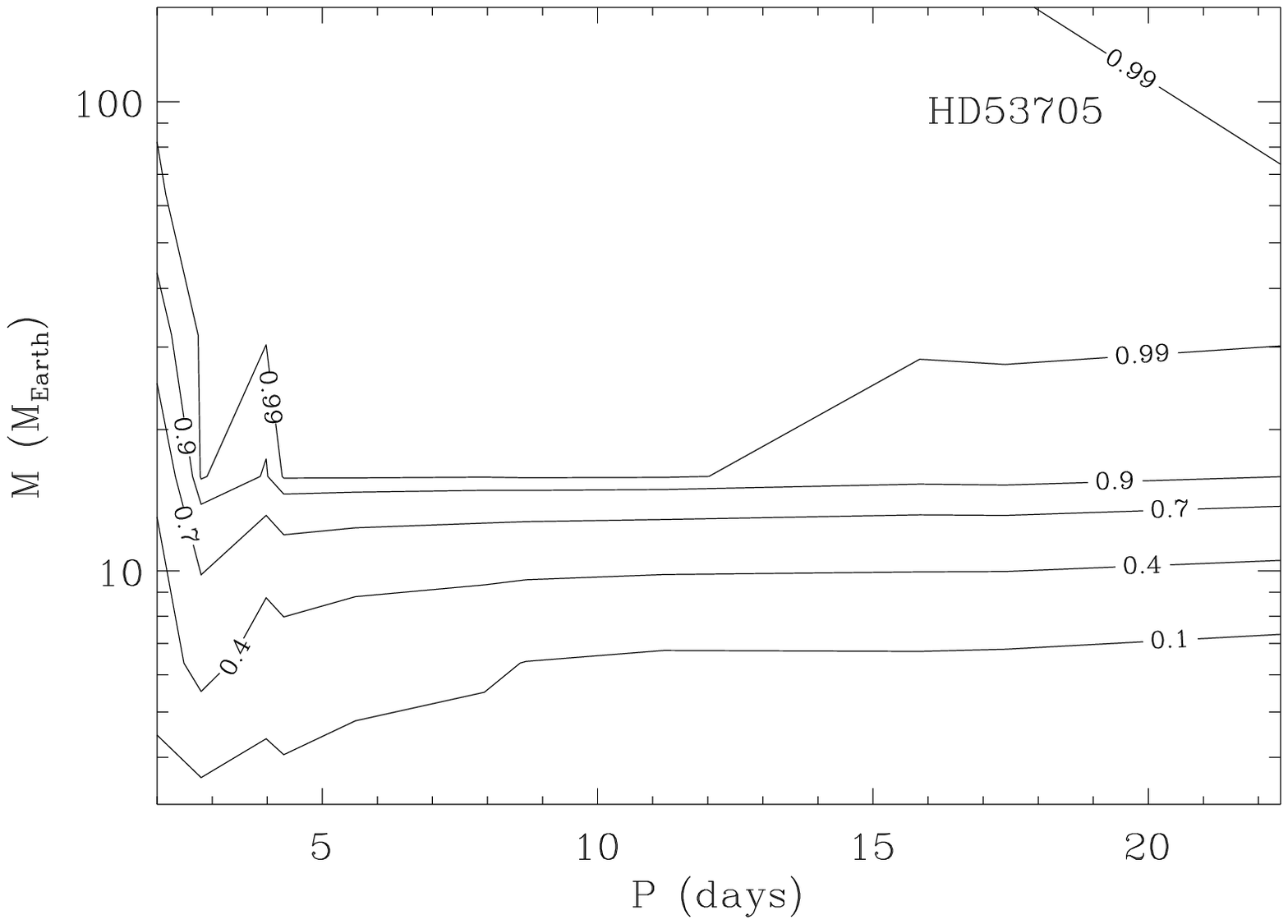}
    \includegraphics[clip=true,width=9.0cm]{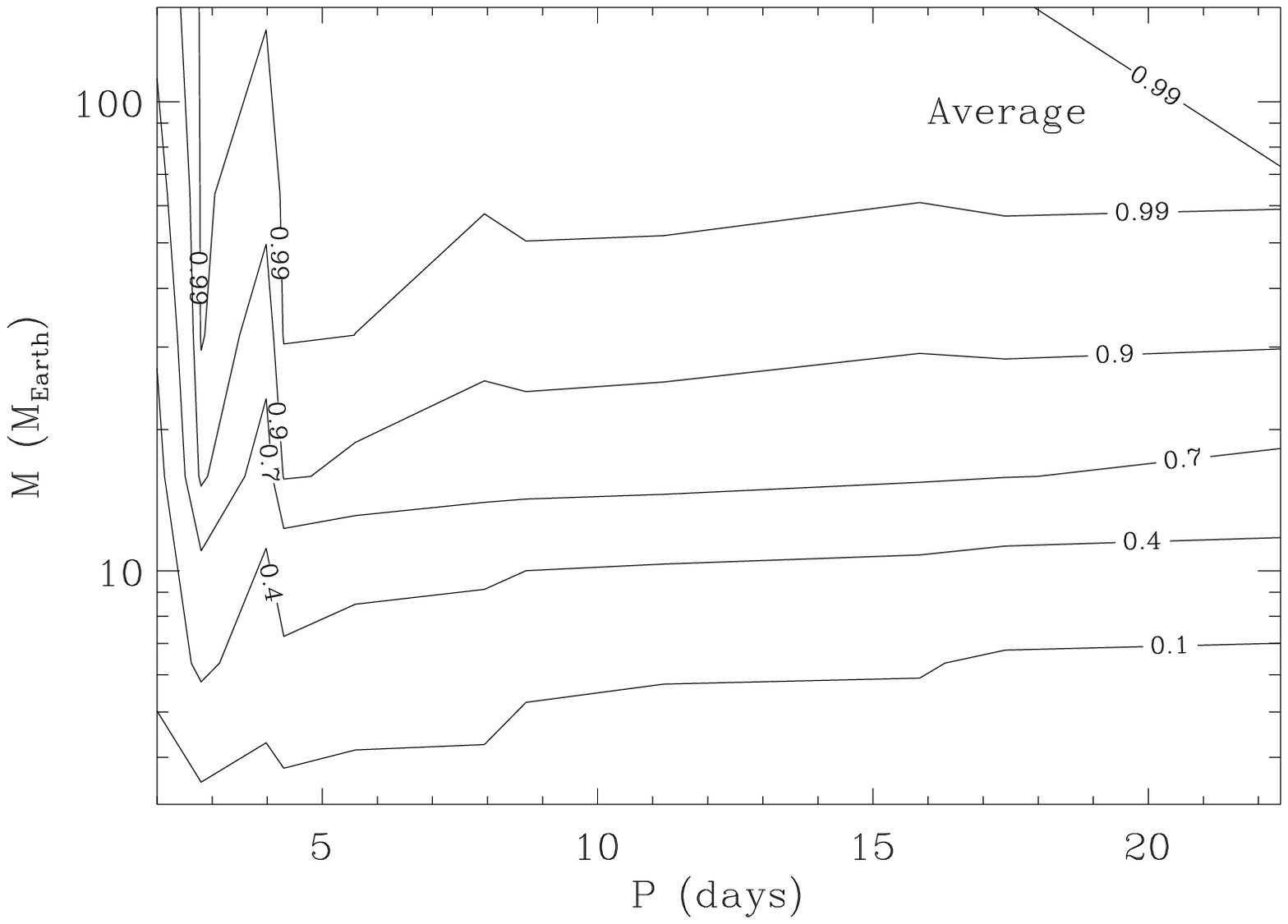}
    \caption{Detectability contours for planets as a function of input
      period and input planet mass (i.e. Doppler \msini\ minimum mass) 
      for two example stars (HD\,4308 and HD\,53705),
      and averaged over all 24 stars. 
      (See text for our formal definition of ``detectability''). The
      known exoplanet  HD\,4308b is shown as an
      asterisk. The data for HD\,4308 allow us to recover
      the previously detected planet -- in comparison  the HD\,53705 data
      demonstrate significantly larger sensitivity to lower-mass planets,
      due to the better sampling that was achieved on this run for this star.
      Averaged over the whole sample, the 10\% detectability contour
      extends down to 3\,\mterr\ for periods from 2-10\,d.
      }
    \label{fig:detect}
  \end{center}
\end{figure}

We can see from the average detectability surface in
Fig. \ref{fig:detect} that the sensitivity of our survey of these 24
stars extends down to detectabilities of 10\% or more at  masses of
5\,\mterr\ over periods of 2-10\,d. For higher mass planets, the
survey is more sensitive, with the detectability being 40\% or better
for all planets with minimum mass $> 14$\,\mterr\ and period $> 3$\,d.

With two planets known with minimum masses of 13.0 and 22.1\,\mterr\
from a survey of 24 stars, the first-order conclusion that can be
drawn from this data set is that for periods of 2-16\,d the frequency
of low-mass (i.e. \msini=10-25\mterr) planets  is $\sim 8 \pm 6$\%.
However, this neglects both the fact that we know that our
detectability is less than perfect (though non-zero) over this mass
range, and the fact that we know that our planet detectability is a
strong function of planet mass and a weak function of planet
period. These effects need to be accounted for if we are to  draw more
robust conclusions from this survey.

For different exoplanet mass functions, we would expect our survey to
have sensitivity to different masses. Put simply, a very steep
exoplanet mass function should counteract our detectability surfaces
to enable the detection of planets in the very lowest masses.
Conversely, a flatter mass function should bias our results toward the
detection of higher-mass planets. So we should be able to use our data
and simulations to provide much more insight into the frequency of
low-mass planets.

If we parameterise the planet mass function as a power-law d$N$/d$M
\propto M^{\alpha}$ (we assume for simplicity that there is no change
in the incidence of planets over this period range at these masses),
we can ask, given our known detectability surface as a function of
mass and period, what normalisation of that power-law would give us
two exoplanet detections from our survey? Given our assumption that
planet frequency is independent of period in this period range, and
because detectability is also a weak function of period in this
range, we make the further simplifying assumption that we can
integrate over the detectability surface at a given mass to derive an
average detectability as a function of mass alone $D_P(M)$.

We can then write the number of planets detected from our survey,
$N_{Det}$, as a function of the number of stars observed $N_{stars}$,
the average frequency of  planets in a given mass and period range,
$F$, the detectability as a function of mass, $D_P(M)$, and the
power-law mass function,

$$N_{Det} = F \times N_{stars} \times \int D_P(M) \times M^{\alpha}\, dM$$

from which (given $N_{Det}=2$) we can derive $F$ (averaged over the
period range P=2-22\,d and the mass range\footnote{Detectability has
been extrapolated to the masses in this range higher than were
simulated, however this contirbutes a negligible uncertainty as the
detectabilities for these more massive mass planets are very close to
1.0} 0.01-20\,\mjup), which varies from 0.14$\pm$0.10 at
$\alpha$=$-0.25$, up to 0.70$\pm$0.85 at $\alpha$=$-1.75$, in the F, G
and K stars probed here.  These planet frequencies might seem high in
comparison to other work, though this is almost certainly due to the
much lower masses  being considered.  For example, in Figure 12 of
\citet{Cumming08}, those authors find an exoplanet frequency of 3\%
for periods less than 16 days and masses above 30\,\mterr. The data
presented here, in comparison, include significant sensitivity down to
4\,\mterr.  The normalisations found here are, of course, uncertain
due to the Poisson-counting statistics of just 2 detections (in a
sample of 24 stars), and do only probe a relatively small range of
periods (or equivalently, semi-major axes).  Having said which, this
survey is important in being one of the first Doppler surveys of
low-mass exoplanets with robustly characterised selection effects.

\begin{figure}
  \begin{center}
    \leavevmode
    \includegraphics[clip=true,width=12.0cm]{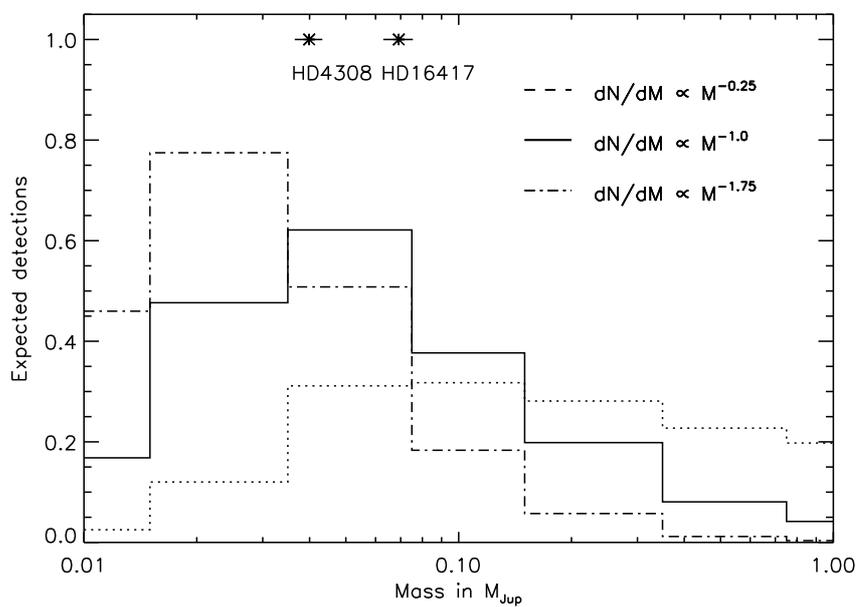}
    \caption{Expected number of exoplanet detections in our survey of 24 stars for 
 different mass functions. The asterisks with uncertainties indicate the 
exoplanets HD\,16417b and HD\,4308b with \msini\ masses of 
0.0692 and 0.0442\,\mjup\ respectively.} 
    \label{fig:mf}
  \end{center}
\end{figure}

Fig. $\ref{fig:mf}$ shows how the  expected number of detected planets
varies as a function of different underlying mass functions (along
with the \msini\ minimum masses for the two detected planets in this
survey,  HD\,4308b and HD\,16417b). As expected, a different peak  in
the expected number of {\em observed} exoplanets is predicted for
different mass functions.  $\alpha$=$-1.75$ leads to a peak in the
number of detected planets around 0.02\,\mjup\ (6.3\,\mterr),  the
$\alpha$=$-1.0$ expected detections peak at around 0.05\,\mjup\
(16\,\mterr) while the flatter $\alpha$=$-0.25$ mass function produces
an expected detection peak at around 0.10\,\mjup\ (32\,\mterr). 

Given that we have found two exoplanet signals in the 16-25\,\mterr\
minimum mass range (corresponding to the 0.05\,\mjup\ or 16\,\mterr\
bin in Fig. \ref{fig:mf})  we may consider these as providing the
first available rough limits on the underlying mass function as
determined by how changes in the mass function cause the peak of
detections  to move toward and away from the 0.05\,\mjup\ (16\,\mterr)
bin. We find the peak moves from the 0.02 to the 0.05\,\mjup\ bin for
$\alpha$ values shallower than $\alpha$=$-1.3$ and from 0.05\,\mjup\
to 0.1 \mjup\ at $\alpha$=$-0.3$ (corresponding to F values of of
0.48$\pm$0.34 and 0.15$\pm$0.10 respectively). Our detections thus
indicate the mass function lies within these limits and that the
exoplanet mass function at low masses  is more likely to be roughly
flat $\alpha \sim -1$ rather than steep.  Although the constraints
from this initial Rocky Planet Search observing run are relatively
modest, our simulations offer a direct methodology to determine an
empirical exoplanet mass function. They can readily be improved with
(1) an expanded sample of high precision data on a larger sample, (2)
simulations covering larger sets of high-precision AAPS matching the
quality obtained in this intensive observing run, (3) higher
resolution simulations (i.e. more realisations at more closely sampled
orbital parameters) and (4) further improvements in our understanding
of stellar jitter and convective granulation.

\acknowledgements

We would like to thank the anonymous referee for comments and suggestions that helped improve the paper. 
We acknowledge support from the following grants:
NSF AST-9988087, 
NASA NAG5-12182, 
PPARC/STFC PP/C000552/1, 
ARC Discovery DP774000;
and travel support from the Carnegie Institution
of Washington and the Anglo-Australian Observatory.  
This research has made use of NASA's
Astrophysics Data System, and the SIMBAD database, operated at CDS,
Strasbourg, France. We are extremely grateful for the uniformly excellent and extraordinary support we 
have received from the
AAT's technical staff over the last 10 years.  
We would also like to extend our effusive thanks to
Professor Matthew Bailes and the
staff at the Swinburne Supercomputing Centre for allowing us the use of
their facilities, and providing support and assistance when required.
The authors acknowledge the use of UCL Research Computing facilities
and services in the completion of this work.

\facility{AAT}

\onecolumn

\clearpage

\clearpage
\begin{deluxetable}{rccclccrrcrrrr}
\rotate
\tablenum{1}
\tablewidth{0pt}
\tabletypesize{\footnotesize}
\tablecaption{Stellar parameters for twenty four Rocky Planet Search targets\label{tab:targets}}
\tablehead{
\colhead{HD} &\colhead{RA}& \colhead{Dec} & \colhead{$V$}  & \colhead{SpTyp} & \colhead{\lrhk}  & \colhead{$P_{\mathrm{max}}$}
                                                             & \multicolumn{2}{c}{[Fe/H]}
                                                                           & \colhead{$v\sin i$}
                                                                                 &\multicolumn{2}{c}{Age}
                                                                                             & \multicolumn{2}{c}{Mass} \\
\colhead{}  &\colhead{}  & \colhead{}     & \colhead{(mag)}& \colhead{} &\colhead{} & \colhead{(s)} &\colhead{} &\colhead{} & \colhead{(\kms)} &\multicolumn{2}{c}{(Gyr)}
                                                                                             & \multicolumn{2}{c}{($M_\odot$)}\\
\colhead{}  &\colhead{}  &\colhead{}      &\colhead{}      & \colhead{} &\colhead{} &\colhead{} & \colhead{VF05} & \colhead{N04}  &\colhead{} & \colhead{VF05} & \colhead{N04}  & \colhead{VF05} & \colhead{N04}  
}
\startdata
1581   & 00 20 04.2 & -64 52 29 & 4.23 & F9.5V & -4.92 & 314 & -0.22&-0.18 & 3.0 &  5.3 &  8.2 & 1.00 & 0.97 \\ 
4308   & 00 44 39.2 & -65 38 58 & 6.55 & G5V   & -5.07 & 234 & -0.31&-0.34 & 0.2 &  9.5 & 17.1 & 0.91 & 0.85 \\ 
10361  & 01 39 47.2 & -56 11 44 & 5.81 & K5V   & -4.88 & 216 & -0.22&-0.17 & 1.9 &  4.5 &\nodata& 0.77 & 0.78 \\ 
10360  & 01 39 47.7 & -56 11 34 & 5.87 & K0V   & -4.95 & 207 & -0.23&-0.16 & 2.2 &  5.5 &\nodata& 0.75 & 0.79 \\ 
10700  & 01 44 04.0 & -15 56 15 & 3.50 & G8V   & -4.94 & 214 & -0.52&-0.42 & 1.3 &\nodata&\nodata& 0.97 & 0.81 \\ 
16417  & 02 36 58.6 & -34 34 41 & 5.79 & G5IV  & -5.08 & 614 & +0.13&+0.01 & 2.1 &  5.8 &  7.6 & 1.18 & 1.1  \\ 
20794  & 03 19 55.6 & -43 04 11 & 4.27 & G8V   & -4.98 & 333 & -0.41&-0.30 & 1.5 & 13.5 &  8.3 & 0.82 & 0.83 \\ 
23249  & 03 43 14.9 & -09 45 55 & 3.52 & K0V   & -5.18 & 883 & +0.16&\nodata& 2.6 &  6.6 &\nodata& 1.19 &\nodata\\ 
26965  & 04 15 17.6 & -07 38 40 & 4.43 & K1V   & -4.87 & 260 & -0.28&-0.05 & 0.5 & 12.2 & 16.9 & 0.78 & 0.84 \\ 
28255A & 04 24 12.2 & -57 04 17 & 6.29 & G4V   & -4.85 & 250 & +0.05&+0.08 & 2.7 &  2.8 &  9.1 & 1.07 & 1.05 \\
43834  & 06 10 14.4 & -74 45 11 & 5.09 & G6V   & -4.94 & 311 & +0.09&+0.10 & 1.7 &  5.4 & 12.8 & 0.98 & 0.91 \\ 
53705  & 07 03 57.3 & -43 36 29 & 5.54 & G3V   & -4.93 & 324 & -0.21&-0.32 & 1.6 &  7.2 & 12.9 & 0.97 & 0.81 \\ 
72673  & 08 32 52.2 & -31 30 10 & 6.38 & K0V   & -4.89 & 229 & -0.37&-0.47 & 0.0 &  8.1 &\nodata& 0.78 & 0.77 \\ 
73524  & 08 37 19.9 & -40 08 52 & 6.53 & G4IV-V& -4.96 & 388 & +0.12&-0.02 & 3.1 &  3.1 &  7.0 & 1.14 & 1.03 \\ 
84117  & 09 42 14.4 & -23 54 56 & 4.93 & G0V   & -4.97 & 452 & -0.07&-0.14 & 5.7 &  3.1 &  4.6 & 1.15 & 1.09 \\ 
100623 & 11 34 29.9 & -32 50 00 & 5.96 & K0V   & -5.07 & 221 & -0.37&-0.51 & 0.7 &  7.8 &\nodata& 0.77 & 0.76 \\ 
102365 & 11 46 31.0 & -40 30 01 & 4.91 & G5V   & -4.95 & 344 & -0.33&-0.29 & 0.7 & 11.0 & 16.1 & 0.86 & 0.86 \\ 
114613 & 13 12 03.1 & -37 48 11 & 4.85 & G3V   & -5.03 & 727 & +0.24&\nodata& 2.4 &  4.9 &\nodata& 1.28 &      \\ 
115617 & 13 18 24.9 & -18 18 31 & 4.74 & G5V   & -4.95 & 314 & +0.05&+0.04 & 2.2 &  6.3 & 12.3 & 0.95 & 0.89 \\ 
122862 & 14 08 27.1 & -74 51 01 & 6.02 & G2.5IV& -4.99 & 680 & -0.13&-0.36 & 2.6 &  5.9 &  8.4 & 1.13 & 1.02 \\ 
128621 & 14 39 35.0 & -60 50 14 & 1.33 & K1V   & -4.92 & 245 & +0.23&+0.15 & 0.9 &  8.0 &\nodata& 0.89 &\nodata\\ 
128620 & 14 39 36.4 & -60 50 02 & -0.0 & G2V   &\nodata& 415 & +0.21&+0.15 & 2.3 &  4.3 &\nodata& 1.12 &\nodata\\ 
136352 & 15 21 48.1 & -48 19 03 & 5.65 & G4V   & -4.91 & 397 & -0.34&-0.32 & 2.0 & 10.9 & 15.9 & 0.89 & 0.85 \\ 
146233 & 16 15 37.1 & -08 22 06 & 5.49 & G1V   & -5.05 & 336 & +0.03&+0.03 & 2.6 &  4.7 &  8.3 & 1.02 & 0.98 \\ 
\enddata
\tablecomments{  Coordinates and magnitudes are based on the Hipparcos catalogue \citep{Perry97}; 
  spectral types are from \citet{Houk82}; 
  activities are from \citet{hsdb96,CaHKI,Jenkins06}; 
  metallicities, isochrone masses and ages are from \citet[][VF05]{VF05} and \citet[][N04]{Nordstrom04}.}
\end{deluxetable}

\clearpage
\begin{deluxetable}{cccccc}
\tablenum{2}
\tablewidth{0pt}
\tabletypesize{\footnotesize}
\tablecaption{Rocky Planet Search Observations\label{tab:obs}}
\tablehead{
\colhead{Target} & \colhead{$N_{\mathrm{obs}}$} & \colhead{RMS} & \colhead{Median} & \colhead{Act.} & \colhead{Osc.} \\
 & &                                                             & \colhead{Unc.}   & \colhead{jitter} & \colhead{jitter} \\
 &                                              & \colhead{(\ms)} & \colhead{(\ms)} & \colhead{(\ms)} & \colhead{(\ms)} 
}
\startdata
1581   &  24 & 1.90 & 0.65 &  4.44 &  0.20 \\
4308   &  24 & 2.24 & 1.09 &  2.17 &  0.12 \\
10360  &  22 & 4.09 & 0.83 &  2.10 &  0.04  \\
10361  &  18 & 3.79 & 0.77 &  2.10 &  0.04  \\
10700  &  24 & 1.80 & 0.57 &  2.10 &  0.10  \\ 
16417  &  24 & 1.57 & 0.78 &  2.19 &  0.35 \\ 
20794  &  28 & 3.06 & 0.52 &  1.80 &  0.12  \\
23249  &  27 & 1.94 & 0.34 &  4.85 &  0.43 \\
26965  &  28 & 3.22 & 0.55 &  1.80 &  0.07 \\
28255A &  28 & 6.12 & 1.10 &  6.20 &  0.08  \\
43834  &  37 & 3.07 & 0.61 &  1.80 &  0.12 \\
53705  &  44 & 3.02 & 1.02 &  2.10 &  0.16  \\
72673  &  42 & 2.51 & 0.98 &  1.80 &  0.07 \\
73524  &  33 & 4.34 & 1.10 &  2.58 &  0.14 \\
84117  &  37 & 5.43 & 1.11 &  3.00 &  0.23  \\
100623 &  34 & 2.87 & 0.94 &  2.10 &  0.05  \\
102365 &  39 & 2.21 & 0.74 &  2.10 &  0.14  \\
114613 &  37 & 3.66 & 0.57 &  3.42 &  0.45 \\
115617 &  41 & 3.92 & 0.61 &  1.80 &  0.14  \\
122862 &  31 & 3.60 & 1.10 &  2.70 &  0.34 \\
128620 &  20 & 1.06 & 0.25 &  1.80 &  0.23  \\
128621 &  31 & 2.28 & 0.40 &  1.80 &  0.12  \\
136352 &  32 & 4.02 & 1.09 &  2.10 &  0.17  \\
146233 &  23 & 3.21 & 0.88 &  2.19 &  0.16 \\
\enddata
\tablecomments{$N_{\mathrm{obs}}$
  indicates the number of nights on which the object was observed. An
  observation on any particular night always comprises at least three
  separate exposures. Also given is the velocity scatter (RMS), median
  velocity uncertainty, activity
  jitter (see text),  and oscillation jitter \citep[from][]{OTJ08}. 
  The latter is based on the average exposure time for each target and is
  given as a guide.}
\end{deluxetable}

\clearpage
\begin{deluxetable}{rr rr}
\tablenum{3}
\tablewidth{0pt}
\tabletypesize{\footnotesize}
\tablecaption{AAT Doppler velocities for HD\,4308. \label{vel4308}}
\tablehead{
\colhead{JD}         & \colhead{Velocity}&\colhead{JD}         & \colhead{Velocity} \\
\colhead{(-2450000)} & \colhead{(\ms)}   &\colhead{(-2450000)} & \colhead{(\ms)}  
}
\startdata
3665.119898 &   -0.3 $\pm$  0.9 & 4120.935790 &   -2.4 $\pm$  1.7 \\
3666.110456 &   -0.6 $\pm$  0.9 & 4125.954730 &    1.7 $\pm$  1.1 \\
3668.136398 &    2.8 $\pm$  0.8 & 4126.924257 &    2.6 $\pm$  1.2 \\
3669.119458 &    0.6 $\pm$  0.9 & 4127.931946 &    3.7 $\pm$  1.1 \\
3670.126654 &    4.5 $\pm$  1.1 & 4128.927984 &    0.5 $\pm$  1.0 \\
3671.147771 &    3.6 $\pm$  1.3 & 4129.930445 &    1.1 $\pm$  1.0 \\
3698.032791 &   -9.0 $\pm$  1.0 & 4130.924765 &   -0.1 $\pm$  0.9 \\
3700.063296 &   -0.7 $\pm$  0.8 & 4131.937199 &   -0.2 $\pm$  1.1 \\
3702.062004 &    1.1 $\pm$  1.3 & 4133.927357 &   -2.3 $\pm$  0.9 \\
3938.327905 &    5.9 $\pm$  1.0 & 4134.931174 &    0.5 $\pm$  1.0 \\
3942.284604 &   -0.1 $\pm$  0.9 & 4135.948345 &   -1.5 $\pm$  0.9 \\
3943.328172 &    2.6 $\pm$  0.8 & 4136.945095 &    3.1 $\pm$  1.0 \\
3944.283860 &   -1.3 $\pm$  0.9 & 4137.977861 &    6.6 $\pm$  1.3 \\
3945.320288 &   -3.8 $\pm$  0.9 & 4140.931894 &    7.5 $\pm$  1.3 \\
3946.318428 &   -4.6 $\pm$  0.9 & 4145.918589 &    5.0 $\pm$  1.2 \\
3947.328379 &   -5.5 $\pm$  1.0 & 4149.900027 &   -4.6 $\pm$  1.1 \\
4008.152191 &   -1.1 $\pm$  1.0 & 4153.907836 &    1.6 $\pm$  1.1 \\
4009.177898 &   -0.0 $\pm$  1.0 & 4154.918636 &    4.5 $\pm$  1.2 \\
4010.168991 &    0.2 $\pm$  0.9 & 4253.327811 &    0.4 $\pm$  0.9 \\
4011.141538 &   -2.4 $\pm$  0.9 & 4255.246671 &   -3.1 $\pm$  1.3 \\
4012.129794 &   -2.5 $\pm$  0.9 & 4256.321705 &   -2.6 $\pm$  1.2 \\
4013.171260 &   -0.2 $\pm$  0.9 & 4257.304964 &   -3.6 $\pm$  1.7 \\
4014.188894 &   -2.6 $\pm$  1.2 & 4334.157294 &    3.0 $\pm$  1.2 \\
4016.210211 &   -1.5 $\pm$  1.0 & 4335.254660 &    7.8 $\pm$  2.7 \\
4018.142500 &   -0.1 $\pm$  1.0 & 4336.264333 &   -3.7 $\pm$  1.3 \\
4037.046606 &    8.5 $\pm$  0.8 & 4338.306202 &   -8.4 $\pm$  1.4 \\
4038.125336 &    4.6 $\pm$  1.0 & 4369.133832 &   -0.3 $\pm$  0.9 \\
4039.054689 &    0.4 $\pm$  1.0 & 4371.241454 &   -1.5 $\pm$  1.0 \\
4040.053384 &   -2.3 $\pm$  0.9 & 4372.138956 &    5.5 $\pm$  1.2 \\
4041.068042 &   -0.4 $\pm$  0.8 & 4373.185170 &    2.7 $\pm$  1.3 \\
4111.941520 &    3.1 $\pm$  1.2 & 4375.180751 &    0.6 $\pm$  1.0 \\
4112.942694 &    4.1 $\pm$  1.1 & 4425.078066 &    0.2 $\pm$  1.1 \\
4113.949634 &    0.3 $\pm$  1.7 & 4429.012543 &   -2.0 $\pm$  1.1 \\
4114.947273 &   -0.4 $\pm$  1.0 & 4429.946086 &   -4.9 $\pm$  0.9 \\
4118.926666 &   -8.9 $\pm$  1.1 & 4431.004964 &   -6.9 $\pm$  1.2 \\
4119.920872 &   -6.7 $\pm$  1.1 & 4431.997618 &   -5.8 $\pm$  1.2 \\
\enddata
\tablecomments{ Julian
Dates (JD) are heliocentric.
Velocities are barycentric but have an arbitrary
zero-point.}
\end{deluxetable}

\begin{deluxetable}{lcc}
\tablenum{4}
\tablecaption{Orbital parameters for HD\,4308b.\label{tab:hd4308}}
\tablewidth{0pt}
\tablehead{
\colhead{Parameter} & \colhead{AAT \& HARPS} & \colhead{HARPS}
}
\startdata
Orbital period $P$ (days)        & 15.609 $\pm$  0.007 &  15.56 $\pm$  0.02   \\
Semi-amplitude $K$ (\ms)         &  3.6   $\pm$  0.3   &  4.07  $\pm$  0.2     \\
Eccentricity $e$                 & 0.27   $\pm$  0.12  &  0.00  $\pm$  0.01     \\
Periastron (JD$-$2450000)        & 108.5  $\pm$  1.9   &3314.7  $\pm$  2.0     \\
$\omega$ (\degr)                 &  210   $\pm$  21    &  359   $\pm$   47    \\
\msini\ (\mterr)                 & 13.0   $\pm$   1.4  &  15.1  $\pm$  0.5\tablenotemark{a}     \\
Semi-major axis (AU)             &  0.118 $\pm$  0.009 &  0.119 $\pm$  0.009\tablenotemark{a}     \\
N$_{\rm fit}$                    &  113                &  41           \\
RMS (\ms)                        &  2.6                &  1.3          \\
$\chi_\nu^2$                     &  1.17               &  1.3          \\
\enddata
\tablenotetext{a}{\msini\ and semi-major axis numbers are those derived from the 
kinematic parameters of \citet{UMB06}, but assuming the same host-star 
mass (0.91$\pm$0.05\,\msun) used in
this paper.}
\end{deluxetable}

\end{document}